\documentclass[11pt]{article}

\usepackage[margin=1in]{geometry}

\usepackage[utf8]{inputenc}
\usepackage{mathtools}
\usepackage{amsmath}
\usepackage{amsthm}
\usepackage{amssymb}
\usepackage{enumerate}
\usepackage{interval}
\usepackage{float}
\intervalconfig{left open fence= (,right open fence=)}
\usepackage[ruled,vlined]{algorithm2e}
\usepackage{standalone}
\usepackage[style=base]{subcaption}
\usepackage{tikz}
\usetikzlibrary{arrows,shapes,positioning,decorations.pathreplacing,calc}
\usepackage{booktabs} 
\usepackage{pgfplotstable} 
\usepackage[normalem]{ulem}
\newcommand{\stkout}[1]{\ifmmode\text{\sout{\ensuremath{#1}}}\else\sout{#1}\fi}
\usepackage{soul}
\usepackage{xcolor}
\pgfplotsset{compat=newest}
\usepackage{hyperref}

\usepackage[section]{placeins}  

\pgfplotstableread[col sep=comma]{data/TUE_server_results_2021_07_12.csv}\allinstances{}
\pgfplotstableread[col sep=comma]{data/CG_summary_2022_08_25.csv}\summary{}
\pgfplotstableread[col sep=comma]{data/TUEserver_summary_opt_2021_07_12.csv}\summaryopt{}

\newtheorem{theorem}{Theorem}

\newtheorem{prop}{Proposition}
\makeatletter
\newcommand\footnoteref[1]{\protected@xdef{}\@thefnmark{\ref{#1}}\@footnotemark}
\makeatother

\usepackage{natbib}
\title{A flow-based formulation for parallel machine scheduling using
	decision diagrams}

\author{Daniel Kowalczyk\thanks{ORSTAT, Faculty of Economics and Business, KU Leuven, Leuven, Belgium; daniel.kowalczyk@kuleuven.be} \and 
Roel Leus\thanks{ORSTAT, Faculty of Economics and Business, KU Leuven, Leuven, Belgium; roel.leus@kuleuven.be} \and
Christopher Hojny\thanks{Combinatorial Optimization Group, Department of Mathematics and Computer Science, Eindhoven University of Technology,  Eindhoven, The Netherlands; c.hojny@tue.nl} \and
Stefan R\o{}pke\thanks{Department of Technology, Management, and Economics, Technical University of Denmark, Lyngby, Denmark; ropke@dtu.dk}}

\date{}
\begin{document}

\maketitle

\begin{abstract}
\footnotesize \noindent We present a new flow-based formulation for identical
	parallel machine scheduling with a regular objective function and without idle time.
	The formulation is constructed with the help of a decision diagram that represents all
	job sequences that respect specific ordering rules.  These rules rely on a partition of the planning horizon into, generally
non-uniform, periods and do not exclude all optimal solutions,
	but they constrain solutions to adhere to a canonical form.
	 The new formulation has numerous variables and constraints,
	and hence we apply a Dantzig-Wolfe decomposition in order to compute the
	linear programming relaxation 
in reasonable
	time; the resulting lower bound is stronger than
	the bound from the classical time-indexed
	formulation. We develop a branch-and-price framework that solves several instances from the literature for the first time.
   We compare the new formulation with the time-indexed and arc-time-indexed formulation by means of a series of computational experiments.
\end{abstract}

\section{Introduction}

We study a scheduling problem 
where  a set $ J = \{1,\ldots,n\} $ of $n$
jobs with processing time~\mbox{\(p_j\in \mathbb{N}\setminus\{0\}\)} for each \(j \in J \) 
needs to be processed by a set  \(M = \{1,\ldots,m\} \) of $m$  identical parallel machines
without pre-emption. 
The problem is to find an assignment of jobs to machines and a sequence
of the jobs on each machine such that some objective function  \(\sum_{j \in
	J}f_{j}(C_{j})\)  is minimized, where $C_j$ is the completion time of $j\in J$.
We focus on parallel machine scheduling with a \textit{regular} objective function, i.e., for which \(f_{j}\) is non-decreasing for all \(j
\in J\). Moreover, we require that there exists an optimal solution without idle
time between the jobs on each machine. Scheduling with weighted completion-time objective
\(Pm||\sum w_j C_j \) is such a problem, where each job~$j$ has a weight \(w_{j}\).  For this objective
the sequencing on each machine is easy (there exists a canonical sequence that can be followed),
so that the difficulty only resides in finding
an optimal division of the jobs over
the machines. Another common scheduling objective 
 is the weighted tardiness, 
where each job \(j\) also has a due date~\(d_{j}\) and
the cost \(f_{j}(C_{j})\) associated with job \(j\) is \(w_{j}T_{j}\)
with \(T_{j} = \max \{0,C_{j} - d_{j}\} \). Since the one-machine case $1|| \sum w_j T_j $ is strongly NP-hard \citep{lawler1977pseudopolynomial}, unless \mbox{P = NP} there is no canonical order of jobs on a machine such that the
weighted-tardiness scheduling problem could reduce to partitioning jobs over the machines, and a different approach is needed.
In the remainder of the paper, we only consider
this weighted-tardiness objective, but the developed method will be generic and other regular objective functions without idle time can
be treated analogously.

The main goal of this paper is to introduce a new flow-based formulation
for  \(Pm||\sum w_j T_j \).
 We 
first discuss some related work in Section~\ref{sec:def}. 
Our formulation is based on a time
discretization of the planning horizon that was introduced
by~\cite{baptiste2009scheduling}, which we summarize in Section~\ref{sec:timehorizon}. 
To the best of our knowledge, we are the first to apply a formulation with a coarser time discretization than the classical time-indexed
formulation (TIF) to a parallel machine scheduling problem.
The formulation itself is presented in  
Section~\ref{sec:form}, and is derived from a binary decision diagram (BDD) that represents all the possible job
sequences on a machine. We show that the LP relaxation of this new integer
linear formulation yields a stronger lower bound than the TIF. 

Our new formulation has many variables and
constraints, which renders 
the computation of the
LP bound inefficient; we apply a Dantzig-Wolfe (DW)
decomposition to resolve this issue.   This reformulation is discussed in Section~\ref{sec:DW}.
We also need to overcome some
convergence problems in the column generation (CG) phase,
which is achieved using the stabilization technique of~\cite{wentges1997weighted} and by variable fixing by reduced cost
as described in~\cite{pessoa2010exact}. 
The running times of the CG phase for the new formulation are
much lower than 
those for the arc-time-indexed formulation (ATIF),
which was introduced independently
by \cite{sourd2009new}, \cite{pessoa2010exact}, and \cite{tanaka2009exact}. At the same time, we find the quality of
the lower bounds from the new formulation to be very similar to that of the ATIF in our experiments.


In Section~\ref{sec:strongbranching} we  develop a branch-and-price (B\&P) algorithm to find optimal integer solutions, 
in which we use an aggressive
strong branching strategy to establish optimality of  primal solutions.
We report on  a series of computational experiments 
in Section~\ref{sec:compexp}, including a comparison with the current state-of-the-art procedure of \cite{oliveira2020improved}.
We conclude the paper in Section~\ref{sec:conclusion}.









\section{Related work}\label{sec:def}

The most popular exact methods for single and parallel machine scheduling use Dynamic Programming (DP), Branch-and-Bound (B\&B)
including Mixed-Integer Programming (MIP) formulations that are
solved by a solver, or a mix of those two techniques. 
The most popular MIP formulations in the literature are  based
on completion variables, (arc-) time-indexed variables, linear ordering
variables, and positional and assignment variables. Below, we discuss
formulations with time-indexed and arc-time-indexed variables.
 For an
extensive introduction to other formulations, we refer
to~\cite{queyranne1994polyhedral}.

\subsection{Time-indexed formulation TIF}
\begin{figure}[t]
	\begin{center}
		\includegraphics[width=0.8\textwidth]{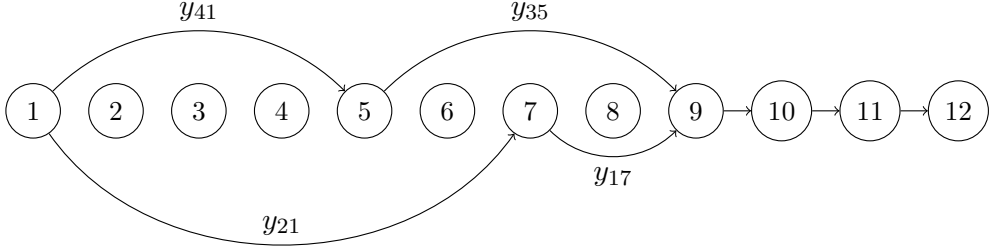}
		\caption{An integral solution to the TIF represented
		as two paths in the DAG}\label{fig:TI}
	\end{center}
\end{figure}

The TIF has been thorougly studied
by, among others, \cite{dyer1990formulating}, \cite{sousa1992time}, and \cite{van1999polyhedral}. With integer
processing times \(p_j\), a sufficiently large
planning horizon \(T\) can be discretized into periods of unit length. Binary
variables \(y_{jt}\) are defined for each job \(j \in J\) and each period \(t
\in \{1,\ldots,T\} \) to decide whether job~\(j\) starts at the beginning of
period \(t\) or not, where period \(t\) starts at time~\(t - 1\) and ends at~\(t\).
The model can be used to represent many different single and parallel
machine scheduling problems (esp.\ with min-sum objective) by adjusting the 
cost parameters
\(\widetilde{c}_{jt}\). 
\begin{subequations}\label{eq:TIform}
	\begin{align}
		\text{minimize }   & \sum_{j \in J}\sum_{t = 1}^{T - p_{j} + 1} \widetilde{c}_{jt}y_{jt} \label{eq:objTI}                                                                                \\
		\text{subject to } & \sum_{t = 1}^{T - p_{j} + 1} y_{jt} = 1                                          & \, & \forall j \in J \label{eq:assTI}                        \\
		                   & \sum_{j \in J}\sum_{s = \max\{ 1, t - p_{j} + 1\}}^{t} y_{js} \le m  & \, & \forall t \in \{1,\ldots,T\} \label{eq:TInummachines}                           \\
		                   & y_{jt} \in \{0,1\}                                                               & \, & \forall j \in J,\, t\in \{1,\ldots,T\} \label{eq:TIvar}
	\end{align}
\end{subequations}
With Constraints~\eqref{eq:assTI} we ensure that every job starts exactly once, while
Constraints~\eqref{eq:TInummachines} impose that at most \(m\) jobs
can be processed in any period. Extra constraints such as release times \(r_{j}\) for
each job \(j \in J\) can be easily modeled by deleting the variables for which
\(t \in \{1,\ldots,r_{j}\} \). 
Solutions to Equations~(\ref{eq:TInummachines}) and~(\ref{eq:TIvar}) 
 can be represented as a flow in a directed acyclic graph (DAG) where the nodes are
associated to the starting period~\(t\) of the jobs and the edges \((t, t +
p_{j})\) are associated to a job \(j\) that starts in period \(t\) and ends in
period \(t + p_{j}-1\). A unit flow from the root node (first period) to the terminal
node (last period) is called a \emph{pseudo-schedule}. A flow satisfies the last
two equations of Formulation~(\ref{eq:TIform}) but not necessarily the
assignment constraints~\eqref{eq:assTI}  and hence there can be pseudo-schedules where two or more edges associated to the same job are
chosen.~In~Figure~\ref{fig:TI} we provide an
optimal integral solution to an instance with \(n = 4\) and \(m = 2\), and with the job data given in Table~\ref{table:instancebdd}. For problem
\(Pm||\sum w_{j}T_{j}\) the time horizon~\(T\) can be chosen as
 \(\lceil (\sum_{j \in J} p_{j} - p_{\max})/m \rceil + p_{\max}\) without
losing all optimal solutions, where \(p_{\max}\) is the maximum processing time
\citep[see, for instance,][]{pessoa2010exact}. In this way we obtain 
\(T = 11\) as a safe upper bound for the time horizon of the instance.

\begin{table}[t]
\centering
	\caption{Job data for the example instance 
\label{table:instancebdd}}{
		\begin{filecontents*}{data/instancebdd.csv}
1,2,4,6
2,6,6,3
3,4,8,2
4,4,8,5
\end{filecontents*}
\pgfplotstabletypeset[
			col sep=comma,
			display columns/0/.style={column name={job \(j\)},string type},
			display columns/1/.style={column name=\(p_{j} \)},
			display columns/2/.style={column name=\(d_{j} \)},
			display columns/3/.style={column name=\(w_{j} \)},
			every head row/.style={before row={\toprule}, after row={\midrule}},
			every last row/.style={after row=\bottomrule},
		]{data/instancebdd.csv} 
	}{}
\end{table}

The TIF is known to have a strong LP bound \citep{dyer1990formulating}, but this strength comes at a cost:
 the length of the planning horizon
is pseudo-polynomial in the size of the instance input, i.e., the number of
constraints and variables depends on the number of jobs and on the processing
times. Hence, the TIF is less applicable for instances with many jobs and large processing times, 
and one cannot always compute the LP relaxation in a
reasonable amount of time. Many specialized techniques have been developed to
overcome this issue; \cite{van2000time}, for instance, use CG to compute the LP relaxation  of the TIF\@. Even using CG, solving the LP relaxation can be slow,
 because the CG phase can suffer from the heading-in effect (when the first iterations of the CG produce irrelevant columns and bad dual
bounds because of the bad dual information), and extreme degeneracy (multiple optimal solutions in the dual and hence the solution of the
restricted master remains constant over several iterations). To cope with this
problem, various approaches were proposed; we refer to \cite{bigras2008time}, \cite{pan2007equivalence}, \cite{sadykov2013column}, and \cite{pessoa2018automation}.
We also note that the TIF can still leave a large duality
gap and hence exact algorithms may need to explore a large B\&B tree. Many polyhedral studies of the TIF were therefore performed, see for
instance \cite{crama1996scheduling}, \cite{sousa1992time}, and \cite{van1999polyhedral}.


\subsection{Arc-time-indexed formulation ATIF}
The ATIF can be seen as an
extended formulation of the TIF\@. The number of variables is a
factor of \(n\) larger than  in TIF\@.
Let \(x_{ij}^{t} \in \{0,1\}\) be variables for
each pair of jobs \(i,j \in J_{+}\), with \(i \neq j\), \(J_{+} =
\{0,1,\ldots,n\} \) and \(p_{0} = 0\), and each \(t \in \{0,\ldots,T\} \). The
variables \(x_{ij}^{t}\) indicate whether or not job \(i\) completes and job \(j\) starts
at time \(t\) on some machine. Let \( \overline{c}_{jt}\) be the cost of starting job \(j\) at time~\(t\) (so $\overline{c}_{jt} = \widetilde{c}_{j,t+1}$).
\begin{subequations}\label{eq:aTIform}
	\begin{align}
		\text{minimize }   & \sum_{i \in J_{+}} \sum_{j \in J\setminus \{i\}} \sum_{t = p_{i}}^{T - p_{j}}  \overline{c}_{jt}x_{ij}^{t} \label{eq:objaTI} &    &                                                    \\
		\text{subject to } & \sum_{i \in J_{+}\setminus \{j\}}\sum_{t = p_{i}}^{T - p_{j}} x_{ij}^{t} = 1                                     & \, & \forall j \in J \label{eq:assaTI}                  \\
		                   & \sum_{\substack{j \in J_{+}\setminus \{i\}                                                                                                                                 \\ t - p_{j} \ge 0}}x_{ji}^{t} -  \sum_{\substack{j \in J_{+}\setminus \{i\} \\ t + p_{i} + p_{j} \le T}} x_{ij}^{t + p_{i}}  = 0  & \, & \forall  i \in J,\,t \in \{0,\ldots,T - p_{i}\} \label{eq:aTIflow} \\
		                   & \sum_{\substack{j \in J_{+}                                                                                                                                                \\ t - p_{j} \ge 0}}x_{j0}^{t} -  \sum_{\substack{j \in J_{+}\\ t + p_{j} + 1 \le T}} x_{0j}^{t + 1}  = 0  & \, & \forall t \in \{0,\ldots,T - 1\} \label{eq:aTIidle}\\
		                   & \sum_{j \in J_{+}} x_{0j}^{0} = m                                                                                & \, & \label{eq:aTIcap}                                  \\
		                   & x_{ij}^{t} \in \mathbb{N}                                                                                        & \, & \forall i \in J_{+},\, j\in J_{+}\setminus \{i\},\nonumber                   \\
		                   &                                                                                                                  &    & t\in \{p_{i},\ldots,T - p_{j}\} \label{eq:aTIvar}  \\
		                   & x_{00}^{t} \in \mathbb{N}                                                                                        & \, & \forall t\in \{0,\ldots,T - 1\} \label{eq:aTIvar0}
	\end{align}
\end{subequations}
Equations~\eqref{eq:aTIflow},~\eqref{eq:aTIidle}, and~\eqref{eq:aTIcap} together with the redundant equation
\begin{equation}
	\sum_{i \in J_{+}} x_{i0}^{T} = m
\end{equation}
model a network flow of \(m\) units over a layered DAG\@.
Since each flow over this network has the same source and destination node, we
can decompose any integral solution into a set of \(m\) paths that correspond to
peudo-schedules.
Constraint~\eqref{eq:aTIidle} models the possibility of idle time in the
solution. With Constraint~\eqref{eq:assaTI} we impose that each job has to be
visited by exactly one path and as a result, each job is assigned to
exactly one machine. \cite{sourd2009new}, \cite{tanaka2009exact}, and \cite{pessoa2010exact}
proposed this formulation independently. \cite{pessoa2010exact} develop a branch-cut-and-price algorithm
to solve the ATIF, so as to handle the
large number of variables, and point out that the ATIF is almost isomorphic to the arc-capacity formulation for the Vehicle
Routing Problem and hence many inequalities for the latter formulation
could be transposed. 
\citeauthor{pessoa2010exact} show that these valid
inequalities can close the gap between a heuristic solution and the LP
bound provided by the relaxation of the ATIF\@.
They also show that the ATIF is
stronger than the TIF, which mainly stems from the fact that direct
repetitions of jobs are forbidden by excluding variables \(x_{jj}^{t}\). One can easily project every solution \(\bar{x}\) of the linear
relaxation of~\eqref{eq:aTIform} onto a solution \(\bar{y}\) of the linear
relaxation of~\eqref{eq:TIform} by setting \(\bar{y}_{jt} = \sum_{i \in
	J_{+}\setminus \{j\} } \bar{x}_{ij}^{t-1}\) for \(j \in J\) and \(t \in \{1,\dots,T - p_{j}+1\} \).
Moreover, simple dominance rules can be applied to the DAG by
omitting the variables \(x_{ij}^{t}\) if permuting jobs \(i\) and~\(j\) at
time~\(t\) decreases the overall cost.  
\cite{oliveira2020improved} continued and refined the work of  \cite{pessoa2010exact}, and their procedure constitutes the current state-of-the-art benchmark for \(Pm||\sum w_j T_j \).

\begin{figure}[t]
	\includegraphics[width=0.8\textwidth]{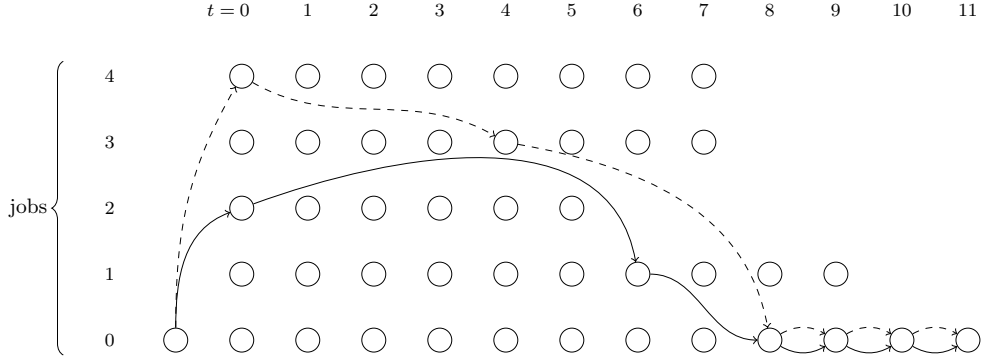}
	\caption{Example of an integral solution of the ATIF represented
		as paths in the DAG, where each edge from job \(i\)
		to job \(j\) that arrives in column \(t\) corresponds to variable
		\(x_{ij}^{t}\)}\label{fig:arcTI}
\end{figure}

An example of the network for the ATIF corresponding to the instance
described in Table~\ref{table:instancebdd} is given in Figure~\ref{fig:arcTI}.
The paths in this solution correspond to schedules \((2,1,0,0,0)\) and
\((4,3,0,0,0)\) on one machine, where each $0$ stands for a unit of idle time. In terms of the variables \(x_{ij}^{t}\) of formulation~\eqref{eq:aTIform},
this solution corresponds with
\(x_{02}^{0}=x_{04}^{0}=x_{43}^{4}=x_{21}^{6}=x_{10}^{8}=x_{30}^{8}=1\),
\(x_{00}^{9} = x_{00}^{10} = x_{00}^{11} = 2\), and the other variables 
equal \(0\).

\section{Discretization of the time horizon}\label{sec:timehorizon}

To cope with the large number of variables in formulations based on a discretization of the time horizon as encountered by TIF and ATIF,
one can resort to a coarser discretization of time.
This is the underlying idea behind the models that were proposed
by~\cite{baptiste2009scheduling} and \cite{boland2016bucket} for single machine
scheduling.

~\cite{boland2016bucket} introduced the bucket-indexed formulation (BIF)\@. Like the TIF, this
formulation partitions the planning horizon into periods of equal lengths but the length of the periods in the BIF is a
parameter, and can be as long as the
processing time of the shortest job. The BIF is equivalent to the TIF if the length of the shortest job is one, but the number
of variables can reduce significantly if the length is larger than \(1\). For a good comparison between the BIF and TIF for single machine scheduling, we refer
to~\cite{boland2016bucket}.

Another formulation that uses a coarser time discretization is the
\textit{interval-indexed model}, which was introduced for single machine scheduling
by~\cite{baptiste2009scheduling}, who partition the planning horizon into time
intervals 
that are defined by a proper
superset of the release dates, due dates, and deadlines of all jobs. \citeauthor{baptiste2009scheduling} show that
 there exists an optimal schedule
where the jobs assigned to a time interval are sequenced according to a modified
weighted shortest-processing-time rule.

We first need the following definition: a \emph{partition} \(\mathcal{I}\) of order~\(q\) of the time horizon is a set of time
intervals \(I_{r}\) given by \(\interval[open left]{e_{r - 1}}{e_{r}}\) with
\(e_{0} = 0\), \(e_{q} = T\), and \(r \in Q = \{1,\ldots,q\} \). We say that a
partition is based on due dates if every \(d_{j}\) equals some~\(e_{r}\),
i.e., \( \{d_{1},\ldots,d_{n}\} \subseteq {\{ e_{r}\}}_{r \in Q} \). A job \(j\) is
assigned to interval \(I_{r}\) if its completion time \(C_{j}\)  is
in~\(I_{r}\). A job \(j\) is on time in interval \(I_{r}\) if \(d_{j} \ge
e_{r}\) and  late if \(d_{j} \le e_{r - 1}\).

Next we define for each interval \(I_{r}\) of \(\mathcal{I}\) a permutation
\(\sigma_{r}\) of \( \{1,\ldots,n\} \), and \(\sigma \) is the set of all
permutations \(\sigma_{r}\) with \(r \in Q\). We say that \(\sigma \) is an \emph{appropriate} set of
permutations for \(\mathcal{I}\) if there exists an optimal schedule in which,
for any interval \(I_{r} \in \mathcal{I}\) and any two jobs \(i,j \in J\)
assigned to the same machine and the same interval~\(I_{r} \), job \(i\) is sequenced
before job~\(j\) when \(\sigma_{r}(i) < \sigma_{r}(j)\).
For the problem \(Pm||\sum
w_j C_j \), for example, there exists a partition of
order~\(1\) with \(I_{1} = \interval[open left]{0}{T}\), where the appropriate
permutation \(\sigma_{1}\) corresponds to Smith's rule, i.e., the jobs are
sequenced in non-increasing order of the ratios \(\frac{w_{j}}{p_{j}}\). Other
problems with a similar priority rule are \(Pm||\sum w_j U_j \) and
\(Pm||\sum w_j V_j \), where function~\(U_{j}\) indicates whether job~\(j\) is late
or not, while \(V_{j} = \min \{p_{j}, \max \{0,C_{j} - d_{j}\} \} \) represents
the portion of work of job~\(j\) that is performed after its due date
\citep[see][]{van1999parallel}.

We say that a partition \(\mathcal{I}\) is \emph{appropriate} if it is possible to
compute an appropriate set of permutations for the partition 
in polynomial time.~\cite{baptiste2009scheduling} show how to find an
appropriate partition for a large number of single machine
problems, and the same approach can be applied to parallel machines
because we need canonical sequences on each machine
separately. We follow \cite{baptiste2009scheduling}
for the construction of an appropriate partition of the time
horizon. Since we consider \(Pm||\sum w_j T_j \), our partition is based on
the due dates \(d_{j}\). Denote by \(\sigma \) a set of permutations for such
a partition; \(\sigma \)~will satisfy an adaptation of
the Weighted Shortest Processing Time (WSPT) and Longest Processing Time (LPT)
rule: 
 first all the late jobs are processed
following WSPT (Smith's) rule and then all the on-time jobs are processed
according to the LPT rule. We also demand that all the late jobs with the
same WSPT ratio be ordered according to the LPT rule. We will make a
distinction between \emph{long} and \emph{short} jobs of an interval \(I_{r}\). A job \(j\)
is long in interval \(I_{r}\) if \(p_{j} \ge e_{r} - e_{r - 1}\), and short 
 if \(p_{j} < e_{r} - e_{r - 1}\). We demand that all long
jobs of interval \(I_{r}\) appear first in \(\sigma_{r}\), meaning
that if \(p_{i} \ge e_{r} - e_{r - 1}\) and \(p_{j} < e_{r} - e_{r- 1}\) then
\(\sigma_{r}(i) < \sigma_{r}(j)\); note that at most one long job can effectively be assigned to $I_r$. For each pair of short jobs \(i,j\) of
interval \(I_{r}\) we require: 

\begin{itemize}
	\item if job \(i\) is late in \(I_{r}\) and job \(j\) is on time in \(I_{r}\) then \(\sigma_{r}(i) < \sigma_{r}(j)\),
	\item if jobs \(i\) and \(j\) are on time and \(p_{i} > p_{j}\) then \(\sigma_{r}(i) < \sigma_{r}(j)\),
	\item if jobs \(i\) and \(j\) are late and \(\frac{p_{i}}{w_{i}} < \frac{p_{j}}{w_{j}}\) then \(\sigma_{r}(i) < \sigma_{r}(j)\),
	\item if jobs \(i\) and \(j\) are late, \(\frac{p_{i}}{w_{i}} = \frac{p_{j}}{w_{j}}\), and \(p_{i} > p_{j}\) then \(\sigma_{r}(i) < \sigma_{r}(j)\).
\end{itemize}

This set of rules was devised in~\cite{baptiste2009scheduling}. The LPT rule  is merely 
a tie-breaker. These rules for each partition are ``almost''
enough to be an appropriate set of permutations for a partition based on due
dates: 
there is
always an optimal solution that satisfies the rules in \(I_{r}\) for each \(r
\in \{1,\ldots,q\} \) except for maybe one job \(j\), and this exception takes
place only if the job \(j\) is late in \(I_{r}\) and is completed first in
\(I_{r}\). 
Based on this observation, the following theorem was derived:

\begin{theorem}[\citealp{baptiste2009scheduling}]\label{th:bapsad}
	A partition \(\mathcal{I} = {\left \{I_{r} \right \}}_{r \in Q}\) is \emph{appropriate} if, for each \(r \in Q\) and each pair of jobs \(i,j \in J\) such that \(\sigma_{r}(i) < \sigma_{r}(j)\), at least one of the following conditions holds:
	\begin{align}
		e_{r}     & \le e_{r - 1} + p_{j}\label{eq:bapsad1}                                                \\
		e_{r - 1} & \ge d_{i} + \left\lceil \frac{w_{j}p_{i}}{w_{i}}\right\rceil - p_{i}\label{eq:bapsad2}
	\end{align}
\end{theorem}

We can start with a partition \(\mathcal{I}\) for which
\( \{d_{1},\ldots,d_{n}\} = \{e_{1},\ldots,e_{q}\} \), i.e., a partition based on
due dates with the smallest number of intervals. By dividing some
intervals, we can obtain a partition that satisfies the conditions of
Theorem~\ref{th:bapsad}. \cite{baptiste2009scheduling} construct an algorithm
that finds such a partition  
for single machine problems. 
\cite{clement2015mixed} later showed that the algorithm provided in \cite{baptiste2009scheduling} provides an appropriate partition but not always a partition with a minimum number of intervals. \cite{clement2015mixed} presents a method to construct an appropriate partition with a minimum number of intervals; we use his procedure in our implementation.

For the instance in Table~\ref{table:instancebdd}, 
an appropriate partition 
is given by \( I_{1} =
\interval[open left]{0}{4}\), \(I_{2} = \interval[open left]{4}{6}\), \(I_{3} =
\interval[open left]{6}{8}\), and \(I_{4} = \interval[open left]{8}{11}\). The set of
permutations \(\sigma~\) is \(\sigma_{1} = (2,3,4,1)\), \( \sigma_{2} =
(2,3,4,1)\), \( \sigma_{3} = (2,3,4,1)\), and \(\sigma_{4} = (4,2,3,1)\). Clearly, none of the intervals contains a ``special'' pair of jobs, i.e., a pair that does not satisfy the requirements of Theorem~\ref{th:bapsad}.

\section{BDD-based formulation for parallel machine scheduling}\label{sec:form}

\cite{baptiste2009scheduling} present a MIP formulation for single machine problems based on
the ideas in the previous section, with binary variables for the assignment of jobs to intervals, which might also be generalized to parallel machines.  In this work we follow a different approach: we will
use the partition of the time horizon described by \citeauthor{baptiste2009scheduling} to develop a new network-flow-based
formulation.
We will construct a binary decision diagram (BDD) over
which at most \(m\) units of flow are pushed; each unit flow from the root node to the
terminal node will represent a pseudo-schedule. 

\subsection{Introduction to BDDs}

BDDs are data structures that allow to represent and manipulate families of
sets that can be linearly ordered. BDDs were introduced
in~\cite{lee1959representation} and~\cite{akers1978binary} as DAGs that are obtained by reducing binary decision trees that
represent 
 a Boolean
function. Recently, decision diagrams have also been used to
solve discrete optimization problems. 
\cite{bergman2016discrete}, for instance, introduce a generic B\&B algorithm for discrete optimization,
 where relaxed BDDs are used to compute
relaxation bounds and restricted BDDs are used to find feasible solutions.
Another relevant example is \cite{cire2013multivalued}, who  use multi-valued decision diagrams to
solve single machine scheduling problems. We refer to \cite{CastroCireBecksurvey} for a survey of recent advances in the use of decision diagrams for
discrete optimization.

Concretely, a BDD \(B\) is a DAG that has two terminal nodes called
\(\mathbf{1}\) and \(\mathbf{0}\). Every non-terminal node \(i\) is
associated to an element v\((i)\) (the label of node \(i\)) of a set and has two
outgoing edges: the high edge, which points to the high child node hi\((i)\),
and the low edge, pointing to the low child node lo\((i)\). There is also
exactly one node that is not a child of any other node in the DAG\@; this node
is the “highest” node in the topological ordering of the DAG and is called the root 
\(\mathbf{r}\). The size of the BDD can be reduced by removing every node whose high
edge points to the terminal node \(\boldsymbol{0}\) and the incoming edges of
the deleted node are connected to the end node of the low edge.

We describe how a subset \(S\) of a ground set \(V\) induces a path \(P_{S}\) from the root
node to \(\boldsymbol{1}\) in a BDD \(B\). We start at the root node of \(B\)
and iteratively choose the next node in the path as follows: if \(a\) is the
current node on the path, then the next node on the path is hi\((a)\) if
v\((a)\) \(\in S\) and lo\((a)\) otherwise. We call the last node along the path
\(P_{S}\) the output of \(S\) on~\(B\), which is denoted by \(B(S)\); clearly
\(B(S)\) is equal to \(\boldsymbol{1}\) or \(\boldsymbol{0}\). We say that \(B\)
accepts \(S\) if \(B(S) = \boldsymbol{1}\), otherwise we say that \(B\) rejects \(S\). A BDD
\(B\) characterizes a family \( \mathcal{F} \subset 2^{V}\) if \(B\) accepts all
the sets in the family \(\mathcal{F}\) and rejects all the sets not in
\(\mathcal{F}\). 
Since we are only interested in paths from the root node to
the terminal node \(\boldsymbol{1}\), without loss of generality, we can
represent the BDDs without terminal node \(\boldsymbol{0}\).

One can construct a BDD 
associated to a family of subsets in
different ways. In this work  we use the efficient and generic recursive framework
of~\cite{iwashita2013efficient}.
Below we show how to define a
restricted family of pseudo-schedules, which is recursively constructed, based on 
the interval-indexed model 
of~\cite{baptiste2009scheduling}.

\subsection{Constructing a BDD that contains all feasible sequences} \label{sec:constructBDD}

Let \(\mathcal{I} = \{I_{1},\ldots,I_{q}\} \) be an appropriate partition of the
time horizon and \(\sigma = \{\sigma_{1},\ldots,\sigma_{q}\} \) the set of
permutations associated to \(\mathcal{I}\). Each \(\sigma_{r}\) imposes an
ordering \(\prec_{r} \) of the jobs in interval~\(I_{r}\), and we write this as follows: 
\( j_{r}^{1}\prec_{r} \ldots \prec_{r} j_{r}^{n} \),
meaning that if job \(j_{r}^{1}\) is assigned to interval \(I_{r}\) then it is
also the first job in interval \(I_{r}\), otherwise the next job that can be
assigned to interval \(I_{r}\) is job \(j_{r}^{2}\), and so on. This leads to an
ordering \(\prec \) over the entire time horizon: \( j_{1}^{1} \prec j_{1}^{2}
\prec \ldots \prec j_{1}^{n} \prec j_{2}^{1}, \ldots \prec j_{2}^{n} \prec
\ldots,j_{q}^{1} \prec \ldots \prec j_{q}^{n} \). Obviously, for
each \(j \in J\) and \(r\in Q\) there is only one \(i \in
\{1,\ldots,n\} \) such that \(j_{r}^{i} = j\).
 For a feasible schedule we need to choose for each job \(j \in J\) exactly one of
its representations (in one of the intervals). We model this using a BDD
to represent suitable subsets of the set
\( \{j_{1}^{1},j_{1}^{2},\ldots,j_{1}^{n},j_{2}^{1},
\ldots,j_{2}^{n},\ldots,j_{q}^{1},\ldots,j_{q}^{n}\} \), while ensuring that  
 each representation \(j_{r}^{i}\) is
completed in the corresponding interval.

\begin{figure}[t]
	\centering
	\caption{A BDD representing all sequences for the instance described in Table \protect\ref{table:instancebdd}.  Solid lines represent high edges, while dotted lines represent low edges.}\label{fig:instancebdd}
    \includegraphics{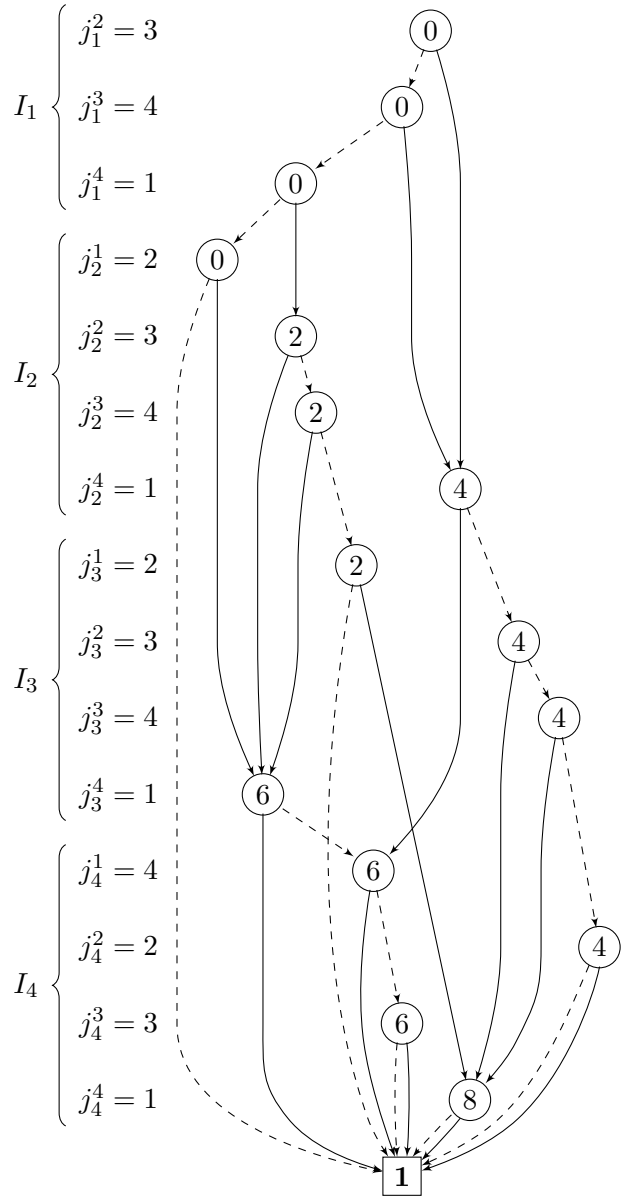}
\end{figure} 

 The BDD will follow the same ordering $\prec$. A configuration
\((j_{r}^{i},t) \) of a non-terminal node is a pair consisting of a
representation \(j_{r}^{i}\) of a job \(j\) that can only be completed in
interval~\(I_{r}\) and the total processing time \(t \) of all the
job representations  
that were chosen before~\(j_{r}^{i}\), so \(t
\) is the starting time of job \(j_{r}^{i} \).
Each node \((j_{r}^{i},t) \) in the BDD apart from the terminal nodes has 
two
child nodes. The high edge, representing inclusion of representation~\(j_{r}^{i}
\), leads to \((j_{r'}^{i'},t + p_j) \), where \(p_{j} \) is the processing time
of the job \(j = j_{r}^{i}\) and \(j_{r'}^{i'}\) is the representation of the
next job \(j'\) that is different from job \(j\) for which  \(t + p_j + p_{j'}
\in I_{r'}  \).   If no such representation~\(j_{r'}^{i'} \) exists, the high edge points to the \mbox{\textbf{1}-node} if
\(t + p_j \in I_{r} \) and to the \textbf{0}-node otherwise.
For the low edge (exclusion of representation \(j_{r}^{i} \)), the same holds but based on the value of \(t \) instead of \(t + p_j \).  


An algorithmic description of the generation procedure of the BDD 
is provided in Algorithm~\ref{alg:childBBTW} in Appendix~\ref{app:BDD}.
Figure~\ref{fig:instancebdd} shows the BDD 
for the instance given in Table~\ref{table:instancebdd}. 
In this instance, representation \(j_{1}^{1}\) can never be
chosen because job \(2\) cannot finish in interval~\(I_{1}\).  It can also be seen that the low edge emanating from $(j^1_2, 0)$, corresponding with the non-selection of job~2 at starting time~0,
immediately leads to the terminal node, because possible next jobs would end too early to complete in intervals $I_2$, $I_3$, or $I_4$, since we do not allow for idle time in the pseudo-schedules.

\subsection{A new flow-based formulation for parallel machine scheduling}

Let \(B = (N, A)\) be the DAG that represents the constructed
BDD\@. 
Each node  \(v\) of the graph is associated to a configuration
\((j_{r}^{i},t)\)
and has two outgoing edges: high edge
\(e_{v}^{1}\) and low edge  \(e_{v}^{0}\). The high edge \(e_{v}^{1}\) 
has a cost \(c_{e_{v}^{1}}  = w_{j}\max \{0,t + p_{j} - d_j\} \) with $j=j_r^i$, while the cost of low edge \(e_{v}^{0}\) is \(0\). The set of all incoming edges of node \(v\) is
given by \(\sigma^{-}(v)\). Let \(A^{0}\) and \(A^{1}\) be the set
of all low and high edges of \(B\), respectively. Let \(p_{B}:A\rightarrow J\) be a map that
projects each edge \(e\) of \(B\) onto the job associated to the head node of
\(e\). 

A formulation for \(Pm||\sum w_j T_j \) can now be constructed using 
a binary variable \(x_{e}\) for each \(e \in A^{1}\) to indicate that
the edge $e$ is chosen, which means that job \(j = p_B(e)\) is completed in interval \(I_{r}\) with completion time \(C_{j} = t +
p_{j}\), where the head node of $e$ has configuration $(p_B(e),t)$.
 For each \(e \in A^{0}\) we define a continuous variable \(x_{e}\) to allow all sequencing decisions to be
represented by a flow from the root node of the BDD to the terminal node
\textbf{1}. The formulation can now be stated as follows:
\begin{subequations}\label{eq:BDDform}
	\begin{align}
		\text{minimize }   & \sum_{e \in A^{1}} c_{e}x_{e} \label{eq:objBDDform}                                                                                 \\
		\text{subject to } & \sum_{e \in A^{1}: p_{B}(e) = j} x_{e} = 1                       & \, & \forall j \in J \label{eq:assBDDform}               \\
		                   & x_{e_{v}^{1}} + x_{e_{v}^{0}} = \sum_{e \in \sigma^{-}(v)} x_{e} & \, & \forall v \in N \setminus \{\textbf{r},\textbf{1}\} \label{eq:flowBDDform} \\
		                   & \sum_{e \in \sigma^{-}(\textbf{1})} x_{e} = m                    & \, & \label{eq:nmachBDDform}                                     \\
		                   & x_{e} \in \{0,1\}                                                & \, &\forall e \in A^{1}\,  \label{eq:var1BDDform}      \\
		                   & x_{e} \ge 0                                                      & \, & \forall e \in A^{0}\,  \label{eq:var2BDDform}
	\end{align}
\end{subequations}

Equations~\eqref{eq:flowBDDform} and \eqref{eq:nmachBDDform} together with
the redundant equation
\begin{equation} x_{e_{\textbf{r}}^{0}} +
	x_{e_{\textbf{r}}^{1}} = m
\end{equation}
can be interpreted as a network flow
of \(m\) units through the BDD from the root node \textbf{r} to the terminal node
\textbf{1}.
Constraints~\eqref{eq:assBDDform} enforce that for each \(j \in J\) we must
choose exactly one edge \(e \in A^{1}\) such that \(p_{B}(e) = j\), meaning that 
we choose exactly one representation of each job across the intervals.  
In what follows, we call the new formulation~\eqref{eq:BDDform} the BDD-based formulation BDDF.

We now show that this formulation will yield better bounds than the TIF~\eqref{eq:TIform}. We show that every solution \(\overline{x}\) of
the LP relaxation of~\eqref{eq:BDDform} can be transformed into a
solution \(\overline{y}\) of the LP relaxation of~\eqref{eq:TIform}.
Consider for this the map \(q_{B}:A^{1} \rightarrow \{1,\ldots,T\}\) that
projects each high edge of the BDD onto the starting period of its head node. 
Let \(\overline{y}_{jt} = \sum_{\substack{e
\in A^{1}:p_{B}(e) = j\\q_{B}(e) = t}}\overline{x}_{e}\) for \(j\in J\) and
\(t \in \{1,\ldots,T - p_{j}+1\}~\). Since \(\overline{x}\) satisfies the assignment
constraints~\eqref{eq:assBDDform}, it follows that \(\overline{y}\) also satisfies the assignment
constraints~\eqref{eq:assTI}, and accordingly 
Constraints~\eqref{eq:flowBDDform} and~\eqref{eq:nmachBDDform} for
\(\overline{x}\) imply Constraint~\eqref{eq:TInummachines} for
\(\overline{y}\).

We now show  that the BDDF can be
strictly better than the TIF for \(Pm||\sum w_j T_j \). Consider the instance described in Table~\ref{table:instancebdd}. An
optimal solution for this instance has cost~\(4\) with the job sequences \((1,4,3)\) and \((2)\) on the two
machines; this integral solution is also an optimal solution to the linear relaxation of the BDDF\@. An optimal solution \(\overline{y}\) of the
LP relaxation of the TIF is  \(\overline{y}_{11} =
\overline{y}_{13} = \overline{y}_{31} = \overline{y}_{37} = 0.5 \),
\(\overline{y}_{21} = \overline{y}_{45} = 1.0\), and all the other variables 
equal to~\(0\). Clearly, this solution to the LP
relaxation of the TIF~\eqref{eq:TIform} cannot be a solution of the
LP relaxation of the BDDF~\eqref{eq:BDDform}, because the
structure of the BDD diagram in Figure~\ref{fig:instancebdd} implies that job
\(1\) can not be assigned to the interval \(I_{1}\) twice. With $\widetilde{c}_{11} = \widetilde{c}_{13} = \widetilde{c}_{21} = \widetilde{c}_{31} = \widetilde{c}_{45} = 0$ and $\widetilde{c}_{37} = 4$, the corresponding objective value equals $2$.  
 We thus obtain:
\begin{prop}
	The BDDF dominates the TIF.
\end{prop}
The ATIF and the new BDDF are not comparable, however; the LP bound of ATIF can be either higher or lower than that of BDDF (see Appendix~\ref{app:notcomparable} for an illustration).

\section{Solving the LP}\label{sec:DW}

\subsection{Dantzig-Wolfe decomposition} \label{subsec:DW}

Formulation~\eqref{eq:BDDform} can have many variables and constraints, which makes a
direct application  restrictive. Following~\cite{van2000time}
and~\cite{pessoa2010exact}, we apply a DW decomposition
to the LP relaxation of Formulation~\eqref{eq:BDDform} (i.e., when all
variables \(x_{e}\) are non-negative reals). This will reduce the number of
constraints from \(|N| + n - 1\) to \(n + 1\), where \(|N|\) is the number of
nodes in the BDD\@. We keep the
assignment constraints~\eqref{eq:assBDDform} and the bounding
constraint~\eqref{eq:nmachBDDform} in the formulation, but we recognize that the 
extreme points of the polytope formed by the flow
constraints~\eqref{eq:flowBDDform} are the paths in the BDD 
from the root node to the terminal \textbf{1}.  Denote the set of all these paths by
\(\mathcal{P}\),  and let \(z_{e}^{p}\) be a parameter that is one if edge \(e\) belongs to
path \(p\) and  zero otherwise.
 We introduce a new variable 
\(\lambda_{p}\) for each \(p\in \mathcal{P}\), with which the LP relaxation of Formulation~\eqref{eq:BDDform} can be re-stated as follows:
\begin{subequations}\label{eq:BDDformlambda}
	\begin{align}
		\text{minimize }   & \sum_{e \in A^{1}} c_{e}x_{e} \label{eq:objBDDformlambda}                                                                     \\
		\text{subject to } & \sum_{e \in A^{1}: p_{B}(e) = j} x_{e} = 1                & \, & \forall j \in J \label{eq:assBDDformlambda}          \\
		                   & \sum_{p \in \mathcal{P}} z_{e}^{p}\lambda_{p} = x_{e}     & \, & \forall e \in A                                      \\
		                   & \sum_{e \in \sigma^{-}(\textbf{1})} x_{e} = m             & \, & \label{eq:nmachBDDformlambda}                                \\
		                   & x_{e} \ge 0                                               & \, & \forall e \in A\,  \label{eq:var2BDDformlambda} \\
		                   & \lambda_{p} \ge 0                                         & \, & \forall p \in \mathcal{P}
	\end{align}
\end{subequations}
Eliminating the variables \(x_{e}\) from
the model, we obtain:
\begin{subequations}\label{eq:BDDLP}
	\begin{align}
		  \text{minimize }  & \sum_{p \in \mathcal{P}}\left(\sum_{e \in A^{1}}c_{e}z_{e}^{p}\right)\lambda_{p} \label{eq:objBDDLP}                                                                         \\
		  \text{subject to }&  \sum_{p\in \mathcal{P}} \left(\sum_{e \in A^{1}: p_{B}(e) = j}z_{e}^{p}\right)\lambda_{p} = 1  &\, & \forall j \in J \label{eq:assBDDLP}                              \\
		 & \sum_{p\in\mathcal{P}} \left( \sum_{e \in \sigma^{-}(\textbf{1})}z_{e}^{p}\right)\lambda_{p} = \sum_{p\in \mathcal{P}}\lambda_{p} = m                              & \, & \label{eq:nmachBDDLP} \\
		 & \lambda_{p} \ge 0  &\, & \forall p \in \mathcal{P}
	\end{align}
\end{subequations}

\subsection{Column generation} \label{subsec:CG}

We solve Formulation~\eqref{eq:BDDLP} with CG, which implies the iterative solution of a \textit{restricted master problem} (RMP), which contains a limited set of columns, and a \textit{pricing
problem}, which checks whether there exists a column with negative reduced
cost. If we assign dual variables \(\pi_{j}\) for \(j \in J\) with
Constraints~\eqref{eq:assBDDLP} and dual variable \(\pi_{0}\) with Constraint~\eqref{eq:nmachBDDLP}, the dual of the LP~\eqref{eq:BDDLP} is given by:
\begin{subequations}\label{eq:BDDdual}
	\begin{align}
		\text{maximize }  & \sum_{j \in J}\pi_{j} + m\pi_{0} \label{eq:BDDobjdual}                                                                                                                            \\
		\text{subject to } & \sum_{j\in J}\left(\sum_{e \in A^1:p_{B}(e) = j}z_{e}^p\right)\pi_{j} + \pi_{0} \leq \sum_{e\in A^1} c_{e}z_{e}^{p} & \, & \forall p \in \mathcal{P}\label{eq:BDDdualpath} \\
		                  & \pi_{j} \in \mathbb{R}                                                                                            & \, & \forall j \in J                                 \\
		                  & \pi_{0} \in \mathbb{R}                                                                                            & \, &
	\end{align}
\end{subequations}
At each iteration of the CG algorithm we check if one of the
constraints~\eqref{eq:BDDdualpath} is violated, meaning that the reduced cost
of the associated column is negative. Recall that a column is a path from the root node to the terminal~\textbf{1} in the BDD that was presented in Section~\ref{sec:timehorizon}.
The pricing problem is then as follows:
given current dual prices \(\overline{\pi}\), can
we find a path \(p \in \mathcal{P}\) such that
\begin{equation}\label{eq:reducedcostBDD}
	\sum_{e\in A^1}
	c_{e}z_{e}^{p} - \sum_{j \in J}\left(\sum_{e \in A^{1}:p_{B}(e) =
	j}z_{e}^p\right)\overline{\pi}_{j} - \overline{\pi}_{0} < 0?
\end{equation}
Inequality~\eqref{eq:reducedcostBDD} can be rewritten as
\begin{equation}
	\left(\sum_{e \in A^1 \cap p}\overline{c}_{e}\right) - \overline{\pi}_{0}  < 0,
\end{equation}
where \(\overline{c}_{e}\) is equal to  \(c_{e} - \overline{\pi}_{p_{B}(e)}\),
which is the reduced cost of high edge \(e\).

A CG algorithm typically needs fewer iterations if one considers constraints
that are strongly violated and hence we will identify paths with the lowest
reduced cost. In this way the pricing problem becomes a
shortest path problem in the BDD, where the length of the high edges \(e\) is
\(\overline{c}_{e}\), and the length of the low edges is zero.
Since the graph is acyclic and has only two
outgoing arcs per node, the running time of a 
labeling algorithm for pricing will be linear in the number of nodes in the BDD \citep{ahuja1993network}.

\subsection{Labeling algorithm} \label{sec:labeling}

In our initial computational experiments we noticed that the pricing algorithm often
generates paths from the root node to the terminal for which the associated
pseudo-schedules contain jobs that are repeated in consecutive positions. The consequence of this
is that the lower bound will tend to be weaker than the
bound from the ATIF (but of course still stronger
than the bound from the TIF\@). In Figure~\ref{fig:instancebdd}, for example, the path  corresponding to the pseudo-schedule \((j_{1}^{4}, j_{2}^{2} , j_{4}^{1} ) = 
( 1 , 3 , 3)\) is allowed, where the path includes the low edge of $(j_3^4,6)$. In Section ~\ref{sec:constructBDD} we mentioned that we avoid two consecutive high edges for the same job, but this does not yet avoid repeated jobs via intermediate low edges.

In principle, one can impose the condition that no job can be visited by a path more than
once: we can compute the intersection of the family of pseudo-schedules and
the family of paths where each job is visited at most once (see \cite{minato1993zero} for 
a generic intersection operation on BDDs).
 In this way, we obtain a BDD that contains exactly all possible schedules. It would be overly time-consuming to construct such a BDD, however, and the pricing problem would also become much harder to
solve because of the number of nodes in the resulting BDD\@. 

It is easier to restrict the pseudo-schedules such that all pairs of consecutive
jobs are different. Note that two jobs assigned to the same time interval will
always be different, so if two successive tasks in a pseudo-schedule
are the same then they are assigned to different intervals. Thus, one might say that the BDDF only ``remembers'' what happens in the same interval. This
implicit memory mechanism is the most fundamental reason why the flow-based formulation
is stronger than the TIF\@.   
We therefore devise a labeling algorithm for pricing that
takes into account that two consecutive jobs in the pseudo-schedule cannot be
the same.   This restriction will have a significant
impact on the quality of the lower bound, and the running time of the
algorithm will still be linear in the number of nodes in the
BDD\@. 


To avoid that two consecutive jobs in the pseudo-schedule are the same, we
need to know the previous job in the optimal path to avoid that the
same job is scheduled consecutively.  We therefore maintain a bucket with two entries 
at each node in the BDD: each entry contains a distance label and the identification of the
previously selected job to achieve that distance label. The first entry is the lowest cost to reach
the node, and the second entry is the lowest cost while not passing via the same predecessor job
as the first entry.  This modified labeling algorithm can be implemented in a forward or backward fashion.
We have observed in preliminary experiments that the forward
labeling algorithm is more time-consuming than the backward variant because the number of label updates is
higher in that case. In the forward labeling algorithm, we may
have to update the labels more often because the in-degree (the number of
incoming edges) of each node can be higher than the out-degree (the number of
outgoing edges), which is at most two. Hence, in our
computational experiments, we use the backward labeling algorithm to find
 paths with minimal reduced cost. The forward labeling algorithm is used
to remove nodes from the BDD \((N, A)\) by reduced cost fixing. A more detailed description of the forward and backward algorithm
is provided in Appendix~\ref{app:label}.

\subsection{Stabilization}
The convergence of the CG algorithm can be slow because of primal
degeneracy. This problem can be circumvented by applying stabilization methods for CG\@.
We apply a smoothing method that was developed by~\cite{wentges1997weighted}, in which we correct the optimal solution of the dual problem of the RMP based on information from the previous iterations
 before plugging it into the
pricing problem. For details of this technique and of stabilization in general, we refer to~\cite{pessoa2018automation}.

\subsection{Reduced cost fixing}

Another method to improve the convergence of the CG is that of 
fixing edges of the BDD\@. 
We can fix the flow on edge \(e\in A^1\) to \(0\) (so remove the high edge from the graph) if
\begin{equation}   LB + (m - 1)\overline{c} +
	\overline{c}_{e} \geq UB,
\end{equation}
where \(\overline{c}_{e}\) is the best reduced cost of a path from the
root node to the terminal node \textbf{1} that traverses the edge
\(e\), \(\overline{c}\) is the reduced cost of the shortest path from
the root node to node~\textbf{1}, \(LB\) is the current
lower bound of the RMP of~\eqref{eq:BDDLP}, and \(UB\) is the best known upper
bound of the optimal cost. In this way, we only remove arcs that will not improve the current best solution.
The computation of  \(\overline{c}_{e}\) uses the forward and backward distance labels discussed in Section~\ref{sec:labeling}.~

Fixing edges by
reduced cost not only has a beneficial effect on the running time of the pricing
algorithm, but can also speed up the B\&B procedure for solving the integer formulation by making the BDD smaller (as a pre-solving step) and via better lower bounds \citep[see][]{irnich2010path,pessoa2010exact}.
Computationally, we find that applying variable fixing each time a number of CG iterations has past, performs better than only doing this at the end of the CG\@.


\section{Branch and price} \label{sec:strongbranching}

In order to find optimal integer solutions for the BDDF, we embed the CG into a B\&B search tree, leading to a B\&P procedure.
In this work we will branch on Generalized Upper Bound (GUB) constraints of the
form \(\sum_{i \in V} x_{i} = 1 \) for some set \(V \) of binaries. The assignment constraints~\eqref{eq:assBDDform} in the BDDF, which
require the selection of one high edge for each \(j \in J\), are clearly of this form.
Branching is based on a subset \(V' \subsetneq V\) for which the solution of
the LP relaxation at a node satisfies \(0 < \sum_{i \in V'} x_{i} < 1\), where we enforce constraint \(\sum_{i \in V'}
x_{i} = 0\) in one child node and constraint \(\sum_{i \in V \setminus V'}
x_{i} = 0\) in the other child node.  This branching scheme is sometimes also called  GUB Dichotomy.
A clear advantage of
branching over GUB constraints instead of branching over individual variables is
that the tree can be more balanced.

In some cases there exists a logical
ordering of the variables in set \(V\) and then the branching method is called SOS\@
branching.  For Constraints~\eqref{eq:assBDDform}, for each \(j \in J\), we can order the edges in 
$A_j = \{ e \in A^{1}: p_{B}(e) = j\}$  in non-decreasing order of the starting time~\(q(e)\).
In the case of SOS branching, an approach to finding an appropriate subset \(V ' \)
was formulated in~\cite{linderoth1999computational} using the concept of ``reference
rows.'' Suppose that \(a_{1} \leq \cdots \leq a_{|V|} \) are
coefficients in a reference row, then a good set for branching is
\[
	V' = \{j \in V\,|\, a_{j} \leq \sum_{\ell \in V} a_{\ell} x_{\ell}^{*}\},
\]
where \(x^{*}\) is a solution of the linear relaxation. In our case we can set
\(a_{e}\) to \(q(e)\) for \(e \in A_{j}\) for every \(j \in J\). Another
possibility is to choose \(a_{e}\) as \(c_{e}\), because the weighted tardiness
objective is a regular function.

There can still remain multiple branching choices,  namely for
every \(j \in J\) we can branch if the
corresponding high edges in \(A_{j}\) are not integral. We apply strong
branching to make good branching decisions. We take a small set of
branching candidates and evaluate the child nodes heuristically by performing a
small number of CG iterations. This first phase produces a ranking, and in this order we fully evaluate the child nodes.
 If for a number of consecutive full evaluations of the child nodes we do not find better bounds, we terminate the full evaluations and branch on the best candidate.



\section{Computational experiments}\label{sec:compexp}

\subsection{Implementation details and instances}

All algorithms have been implemented in the C++ programming language and
compiled with \textsf{gcc} version 11.2.0 with full optimization pack
\textsf{-O3}. We have used and adjusted the
implementation of~\cite{iwashita2013efficient} that can be found on
Github\footnote{\url{https://github.com/kunisura/TdZdd}}  to construct the BDDs\@. All computational experiments were performed
 on one core of a server with Intel Xeon
E5--4610 at 2.4GHz processors  and 64 GB of RAM under a Linux OS\@. All LPs are
solved with Gurobi 9.1.2 using default settings and only one core.  The source code of the procedures can be retrieved from the KU Leuven Gitlab repository.\footnote{\url{https://gitlab.kuleuven.be/u0056096/parallel-machine-bdd}}

 We use the same instances from
the OR-library as \cite{pessoa2010exact} and~\cite{oliveira2020improved}. 
These instances were generated for the single machine problem with weighted
tardiness objective in~\cite{potts1985branch}. There are \(125\)
instances for each \(n \in \{40,50,100\} \). The processing time \(p_{j}\)
for each \(j \in \{1,\ldots,n\} \) was generated from the discrete uniform distribution on the integers in
\(\interval{1}{100}\) and the weight~\(w_{j}\) was generated similarly from  \(\interval{1}{10}\). It was observed that the difficulty of
\(1||\sum w_j T_j \) depends on two
parameters, namely the relative range of due dates  \(RDD\), and
the tardiness factor \(TF \). The due dates are generated from the discrete uniform
distribution on \(\interval{\frac{P(1 - TF - RDD)}{2}}{\frac{P(1 - TF +
		RDD)}{2}}\), where \(P = \sum_{j \in J}p_{j}\) and \(TF,\,RDD \in
\{0.2,0.4,0.6,0.8,1.0\} \). For each \(n \in \{40,50,100\} \) and each pair
\((RDD,TF)\), five instances were constructed. In order to obtain reasonable instances for parallel machine scheduling, \cite{pessoa2010exact} transformed the instances
of~\cite{potts1985branch} by dividing the due dates by the number of machines  \(m\). The processing times \(p_{j}\) and  weights
\(w_{j}\) are kept the same for each \(j \in J\). 
For each pair \((RDD,TF)\) they only retain the first instance; thus there are \(25\) instances for each \(n
\in \{40,50,100\} \) and each \(m \in \{2,4\} \).

\subsection{Comparison of the LP bounds} \label{subsec:LPresults}

 In this section, we will
present computational results of CG for the LP bound computation of
the TIF~\eqref{eq:TIform}, the ATIF~\eqref{eq:aTIform}, and our new formulation BDDF~\eqref{eq:BDDform}. We have implemented a CG algorithm 
for each of these three formulations, with the same enhancements such as stabilization and reduced cost fixing for all three models.  We also incorporate
the pairwise-interchange-based preprocessing derived from Proposition 2 (and 3) of \cite{pessoa2010exact} in the ATIF; a similar interchange argument is implicitly embedded in the BDDF
only within each interval, while this can benefit the ATIF over the entire time horizon.  In this section, ``BDDF'' refers to the formulation with a standard labeling algorithm in the CG phase,
which can generate consecutive repeated jobs, while ``BDDF$_r$'' stands for  CG with the labeling refinement described in Section~\ref{sec:labeling} that avoids identical jobs in consecutive positions
in a pseudo-schedule.


\begin{table}[t]
\centering
\footnotesize	\caption{Size of the graph for TIF, ATIF, BDDF$_r$, and BDDF \label{tbl:sizetw}}{

		\pgfplotstabletypeset[
			columns={n,m,first_size_graph_mean_TimeIndexed,first_size_graph_amax_TimeIndexed,reduction_mean_TimeIndexed,first_size_graph_mean_ArcTimeIndexed,first_size_graph_amax_ArcTimeIndexed,reduction_mean_ArcTimeIndexed,first_size_graph_mean_BddBackwardCycle,first_size_graph_amax_BddBackwardCycle,reduction_mean_BddBackwardCycle,reduction_mean_BddBackward},
			every head row/.style={
					before row={%
							\toprule
							\multicolumn{2}{c}{}&  \multicolumn{3}{c}{TIF}& \multicolumn{3}{c}{ATIF} & \multicolumn{3}{c}{BDDF$_r$}& \multicolumn{1}{c}{BDDF}\\
							\cmidrule(lr){3-5}\cmidrule(lr){6-8}\cmidrule(lr){9-11}\cmidrule(lr){12-12}
						},
					after row={\midrule},
				},
           every last row/.style={after row=\bottomrule},
			columns/n/.style={column type=r,int detect,column name=\textit{n}},
			columns/m/.style={column type=r,int detect,column name=\textit{m}},
			columns/first_size_graph_mean_TimeIndexed/.style={column type=r,fixed,precision=1,zerofill,column name=\emph{avg size}},
			columns/first_size_graph_amax_TimeIndexed/.style={column type=r,fixed,column name=\emph{max size}},
			columns/reduction_mean_TimeIndexed/.style={multiply with=100,column type=r,precision=1,zerofill,fixed,postproc cell content/.append style={  /pgfplots/table/@cell content/.add={}{\%},} ,column name=\emph{avg red}},
			columns/first_size_graph_mean_ArcTimeIndexed/.style={column type=r,precision=1,zerofill,fixed,column name=\emph{avg size}},
			columns/first_size_graph_amax_ArcTimeIndexed/.style={column type=r,fixed,column name=\emph{max size}},
			columns/reduction_mean_ArcTimeIndexed/.style={multiply with=100,column type=r,precision=1,zerofill,fixed,postproc cell content/.append style={  /pgfplots/table/@cell content/.add={}{\%},} ,column name=\emph{avg red}},
		columns/first_size_graph_mean_BddBackward/.style={multiply with=2,column type=r,precision=1,zerofill,fixed,column name=\emph{avg size}},
			columns/first_size_graph_amax_BddBackward/.style={multiply with=2,column type=r,precision=0,fixed,column name=\emph{max size}},
			columns/reduction_mean_BddBackward/.style={multiply with=100,column type=r,precision=1,zerofill,fixed,postproc cell content/.append style={  /pgfplots/table/@cell content/.add={}{\%},} ,column name=\emph{avg red}},
			columns/first_size_graph_mean_BddBackwardCycle/.style={multiply with=2,column type=r,precision=1,zerofill,fixed,column name=\emph{avg size}},
			columns/first_size_graph_amax_BddBackwardCycle/.style={multiply with=2,column type=r,precision=0,fixed,column name=\emph{max size}},
			columns/reduction_mean_BddBackwardCycle/.style={multiply with=100,column type=r,precision=1,zerofill,fixed,postproc cell content/.append style={  /pgfplots/table/@cell content/.add={}{\%},} ,column name=\emph{avg red}}
		]\summary{}
	}{}
\end{table}

The graphs that represent the ATIF and the
BDDF are much larger than those for the TIF\@. The number of edges for ATIF is \(O(n^{2}T)\), 
while this number is \(O(nT)\) for the TIF\@.  Table~\ref{tbl:sizetw} provides some empirical
evidence for this by comparing the average (\emph{avg size}) and maximum
(\emph{max size}) number of edges in all formulations.  The graphs for BDDF$_r$ and BDDF are obviously the same, so those columns are not duplicated.
  We see that the  number of edges in the graph that represents
the~BDDF falls in between the numbers for the other two formulations. 
The column \emph{avg red} presents the average percentage of edges that were removed via
reduced cost fixing by the end of the CG procedure. We observe that the ATIF benefits the most 
from this variable fixing, with around $90\%$ of the high edges removed for all instance classes, followed by the BDDF, and
finally the TIF has the lowest average reduction, but this still amounts to $76.6\%$ at least across the instance classes.
Model BDDF$_r$ is a bit more restrictive than BDDF and benefits slightly more from variable fixing, but the differences are not very large.


In Tables~\ref{tbl:summarytw1} and~\ref{tbl:summarytw2} we report the runtimes 
of the CG algorithms for TIF, ATIF, BDDF$_r$, and BDDF\@.
The columns \emph{avg time}, \emph{max time}, and \# \emph{opt} contain the average and the maximum CPU time of the algorithms (in seconds), and the number
of instances solved at the root node (out of 25), respectively.
  An instance is said to be solved at the root node when the linear relaxation can confirm optimality 
of an initial heuristic solution. The heuristic in our case is a rudimentary iterated local search mechanism  that changes the position of jobs or groups of jobs, or changes their machine allocation.
This heuristic will also produce the starting solution for our B\&P in Section~\ref{subsec:exactresults}.
We find that the average running time of the CG 
for computing the lower bound with BDDF is significantly less than with ATIF, and also that the time needed for TIF, in turn, is 
a lot lower than with BDDF\@. These observations are completely in line with the size of the graphs in which the pricing procedures
are executed, which was reported in Table~\ref{tbl:sizetw}.  Avoiding consecutive identical jobs via labeling is beneficial:  BDDF$_r$ is consistently faster than BDDF\@. 

\begin{table}[t]
\centering
	\caption{Computation time (in seconds) and number of instances solved at the root for the LP relaxation of TIF and ATIF\label{tbl:summarytw1}}{
		\pgfplotstabletypeset[
			columns={n,m,tot_lb_mean_TimeIndexed,tot_lb_amax_TimeIndexed,opt_sum_TimeIndexed,tot_lb_mean_ArcTimeIndexed,tot_lb_amax_ArcTimeIndexed,opt_sum_ArcTimeIndexed},
			every head row/.style={
					before row={%
							\toprule
							\multicolumn{2}{c}{}&  \multicolumn{3}{c}{TIF}& \multicolumn{3}{c}{ATIF}\\
							\cmidrule(lr){3-5}\cmidrule(lr){6-8}
						},
					after row={\midrule},
				},
         every last row/.style={after row=\bottomrule},
 			columns/n/.style={column type=r,int detect,column name=\textit{n}},
			columns/m/.style={column type=r,int detect,column name=\textit{m}},
			columns/tot_lb_mean_TimeIndexed/.style={column type=r,fixed,precision=2,zerofill,column name=\emph{avg time}},
			columns/tot_lb_amax_TimeIndexed/.style={column type=r,precision=2,zerofill,column name=\emph{max time}},
			columns/opt_sum_TimeIndexed/.style={column type=r,fixed,column name=\emph{\# opt}},
			columns/tot_lb_mean_ArcTimeIndexed/.style={column type=r,precision=2,zerofill,fixed,column name=\emph{avg time}},
			columns/tot_lb_amax_ArcTimeIndexed/.style={column type=r,precision=2,zerofill,fixed,column name=\emph{max time}},
			columns/opt_sum_ArcTimeIndexed/.style={column type=r,fixed,column name=\emph{\# opt}}
		]\summary{}
	}{}
\end{table}

\begin{table}[t]
\centering
	\caption{Computation time (in seconds) and number of instances solved at the root for the LP relaxation of BDDF$_r$ and BDDF\label{tbl:summarytw2}}{
		\pgfplotstabletypeset[
			columns={n,m,tot_lb_mean_BddBackwardCycle,tot_lb_amax_BddBackwardCycle,opt_sum_BddBackwardCycle,tot_lb_mean_BddBackward,tot_lb_amax_BddBackward,opt_sum_BddBackward},
			every head row/.style={
					before row={%
							\toprule
							\multicolumn{2}{c}{}&  \multicolumn{3}{c}{BDDF$_r$} & \multicolumn{3}{c}{BDDF}\\
							\cmidrule(lr){3-5}\cmidrule(lr){6-8}
						},
					after row={\midrule},
				},
         every last row/.style={after row=\bottomrule},
 			columns/n/.style={column type=r,int detect,column name=\textit{n}},
			columns/m/.style={column type=r,int detect,column name=\textit{m}},
			columns/tot_lb_mean_BddBackward/.style={column type=r,precision=2,zerofill,fixed,column name=\emph{avg time}},
			columns/tot_lb_amax_BddBackward/.style={column type=r,precision=2,zerofill,fixed,column name=\emph{max time}},
			columns/opt_sum_BddBackward/.style={column type=r,fixed,column name=\emph{\# opt}},
			columns/tot_lb_mean_BddBackwardCycle/.style={column type=r,precision=2,zerofill,fixed,column name=\emph{avg time}},
			columns/tot_lb_amax_BddBackwardCycle/.style={column type=r,precision=2,zerofill,fixed,column name=\emph{max time}},
			columns/opt_sum_BddBackwardCycle/.style={column type=r,fixed,column name=\emph{\# opt}}
		]\summary{}
	}{}
\end{table}


%

The gap between the starting solution and the LP bound
of the different formulations is given in Table~\ref{tbl:summarytw3}.
The pattern that arises here is not as clear-cut as in the previous two tables: we see from Table~\ref{tbl:summarytw3} that,
despite the smaller graphs and the lower runtimes than the ATIF, the BDDF still yields LP bounds that are quite tight, and
very close on average to the ones produced by ATIF\@.  We conclude that while the BDDF is a formulation
that is positioned between the TIF and the ATIF in terms of runtimes and graph size for CG, the LP bounds produced by the BDDF are of rather similar quality as the ATIF, which makes the formulation promising
for finding optimal integer solutions.


\begin{table}[t]
\centering
	\caption{Gap from the starting solution for the formulations TIF, ATIF, BDDF$_r$, and BDDF\label{tbl:summarytw3}}{
		\pgfplotstabletypeset[
			columns={n,m,gap_mean_TimeIndexed,gap_amax_TimeIndexed,gap_mean_ArcTimeIndexed,gap_amax_ArcTimeIndexed,gap_mean_BddBackwardCycle,gap_amax_BddBackwardCycle,gap_mean_BddBackward,gap_amax_BddBackward},
			every head row/.style={
					before row={%
							\toprule
							\multicolumn{2}{c}{}&  \multicolumn{2}{c}{TIF}& \multicolumn{2}{c}{ATIF} & \multicolumn{2}{c}{BDDF$_r$} & \multicolumn{2}{c}{BDDF}\\
							\cmidrule(lr){3-4}\cmidrule(lr){5-6}\cmidrule(lr){7-8}\cmidrule(lr){9-10}
						},
					after row={\midrule},
				},
           every last row/.style={after row=\bottomrule},
			columns/n/.style={column type=r,int detect,column name=\textit{n}},
			columns/m/.style={column type=r,int detect,column name=\textit{m}},
			columns/gap_mean_TimeIndexed/.style={multiply with=100,column type=r,precision=2,zerofill,postproc cell content/.append style={  /pgfplots/table/@cell content/.add={}{\%},}, column name=\emph{avg}},
			columns/gap_amax_TimeIndexed/.style={multiply with=100,column type=r,precision=2,zerofill,postproc cell content/.append style={  /pgfplots/table/@cell content/.add={}{\%},}, column name=\emph{max}},
			columns/gap_mean_ArcTimeIndexed/.style={multiply with=100,column type=r,precision=2,zerofill,postproc cell content/.append style={  /pgfplots/table/@cell content/.add={}{\%},}, fixed,column name=\emph{avg}},
			columns/gap_amax_ArcTimeIndexed/.style={multiply with=100,column type=r,precision=2,zerofill,postproc cell content/.append style={  /pgfplots/table/@cell content/.add={}{\%},}, fixed,column name=\emph{max}},
			columns/gap_mean_BddBackward/.style={multiply with=100,column type=r,precision=2,zerofill,fixed,postproc cell content/.append style={  /pgfplots/table/@cell content/.add={}{\%},}, column name=\emph{avg}},
			columns/gap_amax_BddBackward/.style={multiply with=100,column type=r,precision=2,zerofill,fixed,postproc cell content/.append style={  /pgfplots/table/@cell content/.add={}{\%},}, column name=\emph{max}},
			columns/gap_mean_BddBackwardCycle/.style={multiply with=100,column type=r,precision=2,zerofill,fixed,postproc cell content/.append style={  /pgfplots/table/@cell content/.add={}{\%},}, column name=\emph{avg}},
			columns/gap_amax_BddBackwardCycle/.style={multiply with=100,column type=r,precision=2,zerofill,fixed,postproc cell content/.append style={  /pgfplots/table/@cell content/.add={}{\%},}, column name=\emph{max}}
		]\summary{}

	}{}
\end{table}

\subsection{Comparison of exact procedures} \label{subsec:exactresults}

In this section we present the computational results of the overall B\&P algorithm
based on the BDDF formulation. We compare our algorithm with the currently most competitive procedure in the literature, which is the one by~\cite{oliveira2020improved}.
 In what follows we refer to our new B\&P procedure based on the BDDF simply  as ``BDDF,'' and to the algorithm devised by~\cite{oliveira2020improved} and which is based on ATIF as ``ATIF.''    
We incorporate  the labeling refinement described in Section~\ref{sec:labeling} (which was previously referred to as BDDF$_r$) into BDDF\@.
\cite{oliveira2020improved} feed the solutions found by the heuristic of \cite{DBLP:journals/corr/KramerS15} into their procedure as initial primal bounds, while 
BDDF computes an initial solution using the local search procedure mentioned in Section~\ref{subsec:LPresults}.


The processors in our hardware are clearly slower than those used by \cite{oliveira2020improved}:  the CPU in \cite{oliveira2020improved} has around 30\%
higher clock speed (according to benchmarking websites\footnote{See, for instance, \url{https://www.cpubenchmark.net/} for a comparison of the CPUs.}).  Hence, we transform the results of our
algorithm accordingly, namely we multiply our results by factor~\(0.7\). The time
limit per instance is set to 7200 seconds for our computations (without rescaling).



As a tool for comparing the computational performance of the two procedures, we will use \textit{performance profiles},
which were proposed as a tool for benchmarking optimization software in \cite{dolan2002benchmarking}.
The idea is to compare the methods by the ratio of each method's runtime to the best runtime, per instance. Let $r_{p,s}$ be this ratio
for method~$s$ on instance~$p$, and let $\rho_s(\tau)$ be the probability for method $s$ that a performance ratio~$r_{p,s}$ for a given instance $p$ is within
a factor $\tau \in \mathbb{R}$ of the best possible ratio. The function  $\rho_s$ is then a performance profile, which can be seen as the (cumulative) distribution
function for the performance ratio over all tested instances. In other words, considering  \(\tau \) as the time needed by an
algorithm normalized with respect to the best algorithm, for each value
of~\(\tau \) a performance profile curve reports the fraction of the data set
for which the algorithm is at most \(\tau \) times slower than the best
algorithm. For a more detailed description of performance profile curves
we refer to~\cite{dolan2002benchmarking}.

\begin{figure}[t]
	\caption{Performance profiles over all the instances solved to optimality by both algorithms ($s = $ BDDF, ATIF)}\label{fig:overallpc}
	\centering\includegraphics{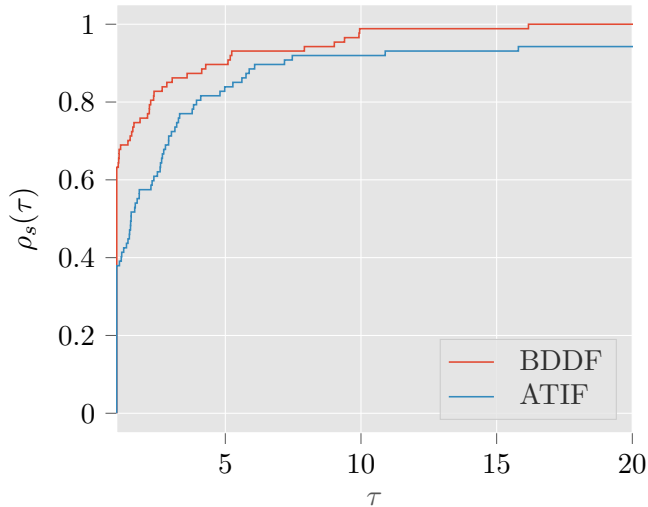}
\end{figure}

In Figure~\ref{fig:overallpc} we plot the performance profiles for BDDF and ATIF for all integer
$\tau = 1,2,3,\ldots$ based on all the instances that were
solved by both methods. Note that \cite{oliveira2020improved} only provide
detailed computational results for instances that
were not ``trivial,'' i.e., not solved in the root node by merely calculating the LP relaxation of the formulation
without additional cuts.  BDDF comes out rather favorably in this plot: from Figure~\ref{fig:overallpc} we can deduce that BDDF is the fastest
algorithm for approximately 65\% of the instances, while this is the case for only almost 40\% for ATIF (this can be read for the entry $\tau =1$ on the horizontal
axis, which is where the plot starts). Algorithm BDDF can solve
approximately 90\% of the  instances within a computing time not exceeding five
times the time for ATIF\@.
Since the details for the trivial instances are not presented 
in~\cite{oliveira2020improved}, we do not fully see the effect of the faster CG
phase of BDDF here. 

\begin{figure}[t]
	\centering
	\caption{Performance profiles per instance class ($s = $ BDDF, ATIF)}\label{fig:per_instance_pc}
	\foreach \i/\j in {2/40, 4/40, 2/50, 4/50,2/100,4/100}{
			\subcaptionbox{\footnotesize{} \(\displaystyle m = \i \) and \(\displaystyle n = \j \)}{\resizebox*{0.45\textwidth}{!}{\includegraphics{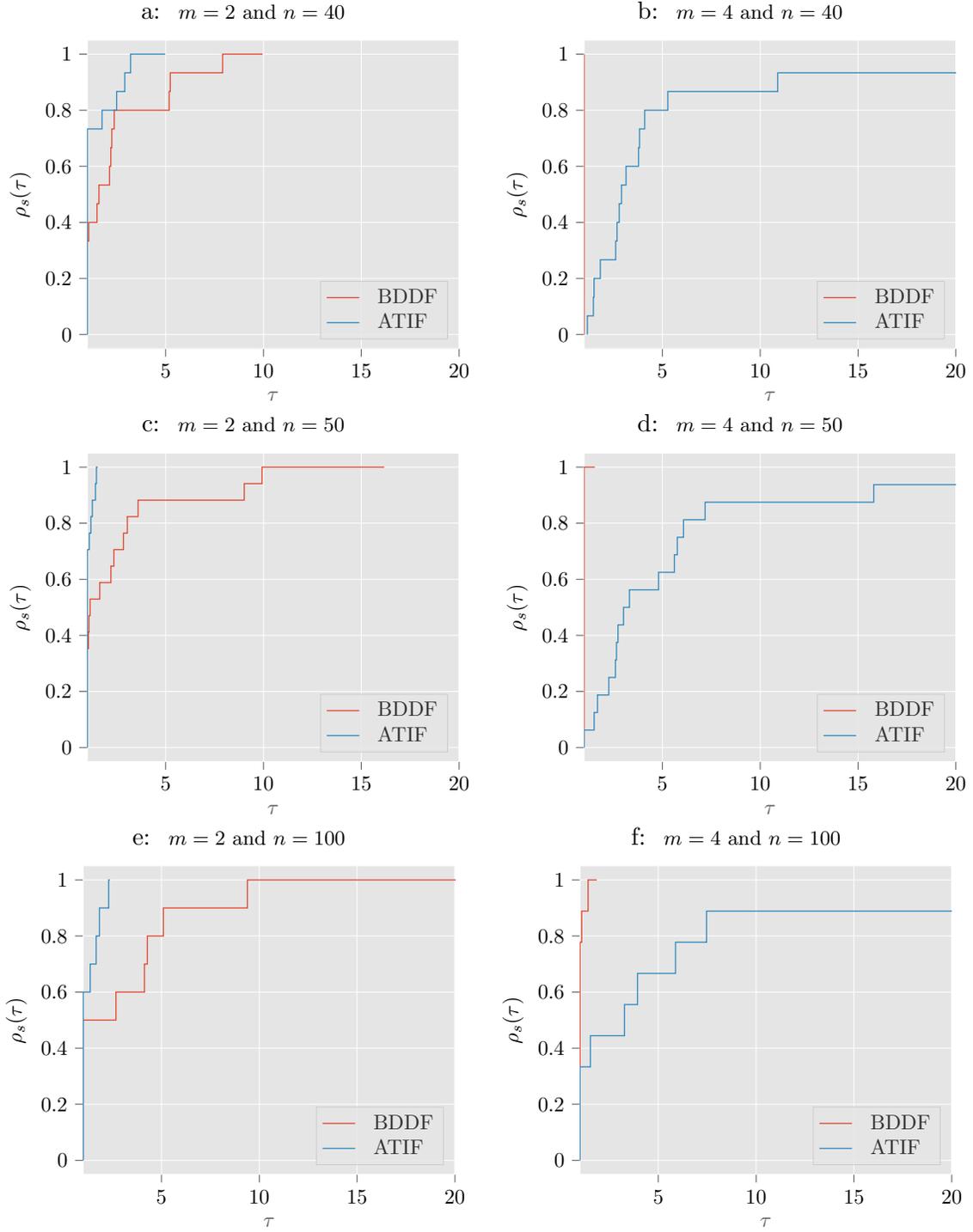}}}
		}
\end{figure}

In Figure~\ref{fig:per_instance_pc} we present performance profiles per instance
class, i.e., per combination \((n, m)\). Clearly, BDDF outperforms ATIF for all instance sets with $m=4$ machines (the three plots on the right side).
Conversely, for instances with two machines (the left plots) ATIF wins the comparison, although the
difference in performance is slightly less pronounced than for the case with four machines.

In Table~\ref{tbl:summarybb} we summarize the results of the two algorithms. Columns \emph{avg time} contain the average running time
over all solved non-trivial instances (in seconds), under \emph{solved} we report the number
of solved instances (out of 25), and in the column  \emph{avg time all} we display the
average time over all solved instances (only for BDDF; in seconds).  Algorithm ATIF solves more
instances to optimality; the runtime limit imposed is not clear, however: some of the instances have taken more than
one day to run in \cite{oliveira2020improved}.  Consequently, a perfect comparison between the two methods is not possible based on this
table. Nevertheless, the overall pattern that was observed in Figure~\ref{fig:per_instance_pc} also occurs here: the runtimes of ATIF
are significantly higher than BDDF for $m=4$, while the differences are not that clear-cut for $m=2$; only for $n=50$ ATIF really dominates BDDF for $m=2$.
We conjecture that the cuts that are used by \cite{oliveira2020improved} are particularly helpful in tightening the formulation especially for instances with few machines.
Overall, for BDDF our strong branching mechanism can close the gap relatively quickly for $n=40$ and $50$, while this is not the case anymore for
instances with \(100\) jobs. In Appendix~\ref{app:details} we present the detailed
computational results of the B\&P procedure based on the formulation BDDF for every instance. We can also report
the optimal solution of three previously unsolved instances in the instance class with \(n = 100\) and
\(m = 4\), namely those with ID number  16, 31, and 56. The only remaining unsolved instance in the data set is the one  with ID number~91 for $n=100$ and $m=2$.

\begin{table}[t]
\centering
	\caption{Summary of the results for the exact procedures\label{tbl:summarybb}}{
		\pgfplotstabletypeset[
			columns={n,m,opt_TimeOliveira_mean_opt_func,OptFound_sum,opt_found_tot_bb_mean_opt_func,opt_sum,opt_tot_bb_mean_opt_func},
			every head row/.style={
					before row={%
							\toprule
							\multicolumn{2}{c}{}&  \multicolumn{2}{c}{ATIF}& \multicolumn{3}{c}{BDDF}\\
							\cmidrule(lr){3-4}\cmidrule(lr){5-7}
						},
					after row={\midrule},
				},
         every last row/.style={after row=\bottomrule},
			columns/n/.style={column type=r,int detect,column name=\textit{n}},
			columns/m/.style={column type=r,int detect,column name=\textit{m}},
			columns/opt_TimeOliveira_mean_opt_func/.style={column type=r,precision=2,zerofill,column name=\emph{avg time}},
			columns/OptFound_sum/.style={column type=r,column name=\emph{solved}},
			columns/opt_found_tot_bb_mean_opt_func/.style={column type=r,precision=2,zerofill,fixed,column name=\emph{avg time}},
			columns/opt_sum/.style={column type=r,fixed,column name=\emph{solved}},
			columns/opt_tot_bb_mean_opt_func/.style={column type=r,precision=2,zerofill,fixed,column name=\emph{avg time all}},
		]\summaryopt{}

	}{}
\end{table}

\section{Conclusion and further research} \label{sec:conclusion}

In this work we have introduced a new formulation for \(Pm||\sum w_j T_j \)
based on binary decision diagrams, which are built using a time discretization
from \cite{baptiste2009scheduling}. We show theoretically and experimentally
that this formulation is stronger than the classical time-indexed informulation, and show
experimentally that this formulation is sometimes weaker and sometimes stronger than the arc-time-indexed formulation.  The computation time of the LP lower bound
of the new formulation with column generation is lower than for the bound computation of the arc-time-indexed model. The reason for this is mainly the size of the
graphs that represent the different formulations.
We have also developed a branch-and-price procedure based on the new formulation; this procedure can solve 
many  instances faster than before,
thanks to strong branching together with the improved running time of the column generation.  Compared with 
the state-of-the-art procedure of \cite{oliveira2020improved}, our new procedure seems to perform better especially
with a larger number of machines.

As a prime avenue for further research, one can examine several techniques from the rich
vehicle routing literature to construct a branch-cut-and-price algorithm for the
new flow-based formulation.  Further closing the gap without branching but rather by introducing cuts
seems to be a logical next step for rendering the resulting algorithm more competitive. \cite{pessoa2010exact} considered such a plan of attack for
the arc-time-indexed formulation for single and parallel machine scheduling, and derived
robust cuts \citep[which do not destroy the structure of the
pricing problem; see also][]{de2003integer}. A similar approach for scheduling on one machine was followed by \cite{van2000time} for the time-indexed formulation.  Potentially, separation for the new formulation could be faster than for the arc-time-indexed model due to its lower number of variables.



A different interesting alternative for continuing this work is to develop a variant of the enumeration algorithm devised in~\cite{baldacci2008exact} for vehicle routing.  The algorithm would  iterate over all the paths from the
	     root node to the  terminal node in the decision diagram with a reduced cost that is
	      less than the duality gap. One can then construct a set-partitioning
	      formulation containing all these paths and hand the resulting formulation to a general
	      MIP solver. In this case, it may be possible to add non-robust cuts to the formulation and to do the pricing by inspection if the number of retained schedules is low enough.

As a final opportunity for further work, one can try to extend the new
flow-based formulation to parallel machine scheduling problems with other
constraints and objective functions. It would be interesting to examine, for
example, whether the formulation can be adapted to the parallel machine
scheduling problem with earliness-tardiness objective, and whether idle time can be incorporated. 

\bibliographystyle{ijocv081}
\bibliography{allpapers-updated} 


\section*{Acknowledgments}
This work was partially 
funded by the European Union's Horizon 2020 research and innovation program under the Marie Skłodowska-Curie grant agreement No.\ 754462.
%

%
%
%


\section*{Appendices}

\setcounter{section}{0}
\renewcommand{\thesubsection}{A.\arabic{subsection}}
\setcounter{table}{0}
\renewcommand{\thetable}{A.\arabic{table}}

\subsection{Generation of the BDD (Section~\ref{sec:constructBDD}) \label{app:BDD}}

Algorithm~\ref{alg:childBBTW} provides a recursive specification of the  decision diagram
that contains all the sequences for a given instance of \(Pm||\sum w_j T_j\), as described in Section~\ref{sec:form}.  The root of 
the BDD has configuration \((j_{1}^{1},0) \), and the function CHILD takes as input a configuration $(j,t)$ of a
node and $b \in \{0, 1\}$ and outputs the configuration of the $b$-child of $(j, t)$, where $0$ and $1$
refer to the low and high edge, respectively.   The terminal nodes \textbf{1} and~\textbf{0} are represented respectively by \((nq + 1, 1) \) and \((nq + 1,0) \).

\begin{algorithm}[h]
	\caption{Recursive specification of the BDD\label{alg:childBBTW}}
	\SetKwProg{Fn}{Function}{}{end}
	\Fn{\(CHILD((j_{r}^{i},t),b) \)}{
		\If{\(b = 1 \)}{
			\(t' \leftarrow t + p_{j_{r}^{i}} \) \;
		}
		\Else{
			\(t' \leftarrow t\)
		}

		\(j_{r'}^{i'} \leftarrow MINJOB(j_{r}^{i},t') \) \;
		\If{\(j_{r'}^{i'} = nq + 1 \)}{
			\If{\(t'  \in I_{r} \)}{
				\Return{}~\( (nq + 1, 1)\) \;
			}
			\Return{}~\( (nq + 1, 0)\)
		}

		\Return{}~\((j_{r'}^{i'},t') \) \;
	}
	\Fn{\(MINJOB(j_{r}^{i},t ) \)}{
		\If{\(\min \{j_{r'}^{i'} \succ j_{r}^{i} | t + p_{j_{r'}^{i}} \in I_{r'}\} \) exists}{
			\Return{}~\(\min \{j_{r'}^{i'} \succ j_{r}^{i} | t + p_{j_{r'}^{i'}} \in I_{r'}\} \) \;
		}
		\Return{}~\(nq + 1 \) \;
	}
\end{algorithm}

\subsection{The ATIF and the BDDF are not comparable} \label{app:notcomparable}


We provide an example instance that shows
that the polyhedron that represents the solution space of the linear relaxation of the formulation ATIF
is not included in the polyhedron of the BDDF\@.
 Table~\ref{table:bdd_tighter} contains the job data for 
the instance with $n = 7$ jobs, and we work with $m=2$ machines.
The lower bound provided by the relaxation of the BDDF is \(117.333\dots\), while the bound for ATIF is \(116.6777\dots\)
In our experiments discussed in Section~\ref{subsec:LPresults} we encountered a number of instances where the LP relaxation of the ATIF provides a tighter bound than the BDDF; for brevity, 
we do not include such an instance here.

\begin{table}[h]
\centering
	\caption{Job data for the example instance 
                  \label{table:bdd_tighter}}{
		\begin{filecontents*}{data/bdd_tighter.csv}
1,92,197,5
2,30,114,6
3,47,86,6
4,19,155,1
5,65,136,5
6,78,158,6
7,82,95,1
\end{filecontents*}
\pgfplotstabletypeset[
			col sep=comma,
			display columns/0/.style={column name={job $j$},string type},
			display columns/1/.style={column name=\(p_{j} \)},
			display columns/2/.style={column name=\(d_{j} \)},
			display columns/3/.style={column name=\(w_{j} \)},
			every head row/.style={before row={\toprule}, after row={\midrule}},
			every last row/.style={after row=\bottomrule},
		]{data/bdd_tighter.csv} 
	}{}
\end{table}

\subsection{Labeling algorithm (Section~\ref{sec:labeling})} \label{app:label}

 For each node \(v\) in the
BDD \((N, A)\) we define a bucket \(F_{v}\) that stores distance labels, representing lengths of partial
paths that end in \(v\). Unlike  traditional labeling algorithms for
shortest-path problems with resource constraints, we store only two labels
in the bucket \(F_{v}\) for each node \(v \in N\), namely the label
\(L^{1}_{v} = (\overline{c}^{1}_{v}, v, pred^{1}_{v})\) for a  partial path \(P^{1}\) leading to $v$ with the best
reduced cost $\overline{c}^{1}_{v}$, and the label \(L_{v}^{2} = (\overline{c}^{2}_{v}, v,
pred^{2}_{v} )\) for a partial path \(P^{2}\) with best reduced cost such
that \(j_{w^{1}} \neq j_{w^{2}}\), 
where $pred^{1}_{v} =  nil$ or contains a pointer to the label of the  predecessor configuration $ w^{1} =  (j_{w^{1}}, t_{w^{1}})$ of $v$ in path $P^1$,
and similarly \( pred^{2}_{v} = nil\) or a pointer to the label of predecessor $w^2 =   (j_{w^{2}}, t_{w^{2}})$ in path $P^2$.
 The corresponding forward recursion in Algorithm~\ref{alg:solvezdd} finds a path \(P \in
\mathcal{P}\) with minimum reduced cost such that all pairs of consecutive
jobs on the machine are different.
After the labeling algorithm, the bucket \(F_{\textbf{1}}\) associated with
the terminal node \(\textbf{1}\) will hold two labels with paths from the root
to the terminal node  \(\textbf{1}\) with the smallest reduced cost and for
which the associated pseudo-schedule is such that 
consecutive jobs are different. We can retrieve these
pseudo-schedules by a simple backtracking algorithm via the pointers
associated to the labels. 

\begin{algorithm}[p]
	\caption{Forward labeling algorithm for pricing\label{alg:solvezdd}}
	\KwData{BDD \((N, A)\), optimal solution \(\pi\) of the dual program}
	\KwResult{Pseudo-schedule \(s\) with minimum negative reduced cost} 
	\(L_{v}^{1} \leftarrow (\infty, v, nil)\) and  \(L_{v}^{2} \leftarrow (\infty, v, nil),\, \forall v \in N\setminus \{\textbf{r}\} \) \;
	\(L_{\textbf{r}}^{1} \leftarrow (-\pi_{0}, \textbf{r}, nil)\) and  \(L_{\textbf{r}}^{2} \leftarrow (\infty, \textbf{r}, nil)\) \;

	\For{\(v  \in N\setminus \{\bf{1}\} \), with $v =  (j_{v}, t_{v})$, in breadth-first order}{
		Let \(lo(v)\) be the low child of \(v\) and \(hi(v)\) be the high child of \(v\) \;
		\(L_{v} \leftarrow \, \) best label in \(F_{v}\) for which the predecessor job \(j_{w}\) is different from \(j_{v}\) \;
		Extend label \(L_{v}\) to a label in \(F_{hi(v)}\) if the reduced cost of the new label is smaller than the current reduced cost of the label, and the labels are constructed in such a way that the predecessors point to nodes associated with different jobs \;
		Extend the labels \(L_{v}^{1}\) and \(L_{v}^{2}\) to labels in \(F_{lo(v)}\) if the reduced cost of the new labels is smaller than the reduced cost of the current labels in \(F_{lo(v)}\) \;
	}
\end{algorithm}

\begin{algorithm}[p]
	\caption{Backward labeling algorithm for pricing\label{alg:solvezdd2}}
	\KwData{BDD \((N, A)\), optimal solution \(\pi\) of the dual program}
	\KwResult{Pseudo-schedule \(s\) with minimum negative reduced cost} 
	\(L_{v}^{1} \leftarrow (\infty, v, nil)\) and  \(L_{v}^{2} \leftarrow (\infty, v, nil),\, \forall v \in N\setminus \{\textbf{1}\} \) \;
	\(L_{\textbf{1}}^{1} \leftarrow (-\pi_{0}, \textbf{1}, nil)\) and  \(L_{\textbf{1}}^{2} \leftarrow (\infty, \textbf{1}, nil)\) \;

	\For{\(v \in N\setminus \{\bf{1}\} \), with $v = (j_{v}, t_{v})$,  in reversed breadth-first order}{
		First extend the labels of \(B_{lo(v)}\) to labels in \(B_{v}\) \;
		Let \(L_{v}\) be a candidate label of \(B_{v}\) that is extended from one of labels of \(B_{hi(v)}\) such that \(j_{v}\) is different from the job associated to the pointer of label of one of the labels in \(hi(v)\) \;
		If the reduced cost of label \(L_{v}\) is smaller than the current reduced costs of the labels in \(B_{v}\) then update the labels in \(B_{v}\) appropriately \;

	}
\end{algorithm}

As mentioned in Section~\ref{sec:labeling}, the recursion can alternatively be conducted backwards.
In this case the labels will have the structure  \((\widetilde{c}_{v}, v, 
prev_{v})\), with $prev_v = nil$ or a pointer to the previously chosen node (successor node) as part of the partial path, and two labels are stored in a bucket \(B_{v}\) for each node $v$, one for the best
partial path, and one for a path with the best length but with different successor than the first label.
The backwards labeling algorithm is described in pseudo-code 
in Algorithm~\ref{alg:solvezdd2}.


\newpage

\subsection{Detailed computational results for the B\&P procedure based on BDDF} \label{app:details}

In the following tables (Table \ref{tbl:twa} to \ref{tbl:twf}) we include the detailed computational results of our  B\&P procedure based on BDDF for each tested instance.  An instance is solved to guaranteed optimality if and only if \mbox{\emph{LB} = \emph{UB}}\@. The information provided in the different columns is as follows:
\begin{itemize}
\item \#\emph{id} = instance ID number
\item \emph{UB} = best found upper bound (best solution)
\item \emph{LB root} = lower bound in the root node of the B\&B search tree
\item \emph{LB} = best found (global) lower bound
\item \#\emph{iter} = total number of CG iterations across all nodes in the search tree
\item \#\emph{iter root} =number of CG iterations in the root node
\item \#\emph{nodes}  = number of nodes in the B\&B search tree
\item \emph{time LP} = CPU time (in seconds) for solving the LPs in all nodes
\item \emph{time LP root} = CPU time (in seconds) for solving the LP in the root node
\item \emph{time} = total CPU time (in seconds)
\end{itemize}
The CPU times were transformed to be comparable with those obtained by the machine used by \cite{oliveira2020improved}, where 7200 seconds on our computer 
equates with approximately 5040 seconds on the computer of  \citeauthor{oliveira2020improved}.


\foreach\i/\j/\k in {2/40/a,4/40/b,2/50/c,4/50/d,2/100/e,4/100/f}{
                               \begin{table}[h]
                                               \centering
                                               \caption{B\&P results per instance for \(m = \i \) and \(n = \j \)}\label{tbl:tw\k}
                                               \pgfplotstabletypeset[
                                               columns={Inst,global_upper_bound_AFBC,global_lowerbound_root_AFBC,global_lower_bound_AFBC,nb_generated_col_AFBC,nb_generated_col_root_AFBC,nb_nodes_explored_AFBC,tot_lb_AFBC,tot_lb_root_AFBC,tot_bb_AFBC},
                                                               every head row/.style={
                                                                                              before row={%
                                                                                                                              \toprule
                                                                                                              },
                                                                                              after row={\midrule},
                                                                               },
                                               every last row/.style={after row=\bottomrule},
                                                               columns/Inst/.style={column type=r,int detect,column name=\#\emph{id}},
                                                               columns/global_upper_bound_AFBC/.style={column type=r,int detect,column name=\emph{UB}},
                                                               columns/global_lowerbound_root_AFBC/.style={column type=r,int detect,column name=\emph{LB root}},
                                                               columns/global_lower_bound_AFBC/.style={column type=r,int detect,column name=\emph{LB}},
                                                               columns/nb_generated_col_root_AFBC/.style={column type=r,int detect,column name=\#\emph{iter root}},
                                                               columns/nb_nodes_explored_AFBC/.style={column type=r,int detect,column name=\#\emph{nodes}},
                                                               columns/nb_generated_col_AFBC/.style={column type=r,int detect,column name=\#\emph{iter}},
                                                               columns/tot_lb_AFBC/.style={multiply with=0.7,column type=r,fixed,precision=2,zerofill,column name=\emph{time LP}},
                                                               columns/tot_lb_root_AFBC/.style={multiply with=0.7,column type=r,fixed,precision=2,zerofill,column name=\emph{time LP root}},
                                                               columns/tot_bb_AFBC/.style={multiply with=0.7,column type=r,fixed,precision=2,zerofill,column name=\emph{time}},
                                                               row predicate/.code={
                                                                                              \pgfplotstablegetelem{##1}{n}\of{\allinstances}
                                                                                              \pgfmathsetmacro{\xn}{\pgfplotsretval}
                                                                                              \ifnum\xn=\j
                                                                                                              \pgfplotstablegetelem{##1}{m}\of{\allinstances}
                                                                                                              \pgfmathsetmacro{\yn}{\pgfplotsretval}
                                                                                                              \ifnum\yn=\i
                                                                                                              \else
                                                                                                                              \pgfplotstableuserowfalse{}
                                                                                                              \fi
                                                                                              \else
                                                                                                              \pgfplotstableuserowfalse{}
                                                                                              \fi
                                                                               }
                                               ]\allinstances{}

                               \end{table}
                }


\end{document}